\documentclass[preprint,12pt,3p]{elsarticle}

\usepackage{amssymb}
\usepackage[utf8]{inputenc}
\usepackage{amsmath}
\usepackage{amssymb}
\usepackage{tcolorbox}
\usepackage{rotating}
\usepackage{multirow}
\usepackage{footnote}
\usepackage{subfigure}
\usepackage{hhline}
\usepackage{bm}
\usepackage{array,booktabs}
\usepackage{algorithm}
\usepackage{algorithmic}
\usepackage{array}
\usepackage{hyperref}
\usepackage{tabulary}
\usepackage{mdwmath}
\usepackage{mdwtab}

\begin{document}

\begin{frontmatter}

\title{Targeted Advertising Based on Browsing History}

\author[label1]{Yong Zhang\corref{cor1}}
\cortext[cor1]{Corresponding author. The work is done when Yong Zhang worked as an intern in GroupM.}
\ead{yzhang067@e.ntu.edu.sg}
\author[label2]{Hongming Zhou\corref{cor2}}
\ead{hongming.zhou@groupm.com}

\author[label2]{Nganmeng Tan}
\ead{nganmeng.tan@groupm.com}
\author[label2]{Saeed Bagheri}
\ead{saeed.bagheri@groupm.com}
\author[label1]{Meng Joo Er}
\ead{emjer@ntu.edu.sg}

\address[label1]{School of Electrical and Electronic Engineering, Nanyang Technological University, Singapore}
\address[label2]{GroupM, Singapore}

\begin{abstract}
Audience interest, demography, purchase behavior and other possible classifications are extremely important factors to be carefully studied in a targeting campaign. This information can help advertisers and publishers deliver advertisements to the right audience group. However, it is not easy to collect such information, especially for the online audience with whom we have limited interaction and minimum deterministic knowledge. In this paper, we propose a predictive framework that can estimate online audience demographic attributes based on their browsing histories. Under the proposed framework, first, we retrieve the content of the websites visited by audience, and represent the content as website feature vectors; second, we aggregate the vectors of websites that audience have visited and arrive at feature vectors representing the users; finally, the support vector machine is exploited to predict the audience demographic attributes. The key to achieving good prediction performance is preparing representative features of the audience. Word Embedding, a widely used technique in natural language processing tasks, together with term frequency-inverse document frequency weighting scheme is used in the proposed method. This new representation approach is unsupervised and very easy to implement. The experimental results demonstrate that the new audience feature representation method is more powerful than existing baseline methods, leading to a great improvement in prediction accuracy.
\end{abstract}


\end{frontmatter}


\section{Introduction}
Nowadays people spend a great amount of time online doing all kinds of things, like reading news, playing games, shopping, etc. This consumer behavior results in advertisers putting great efforts and investment in online advertising. However, advertising audiences or web users, in this case, seldom click on online ads, which significantly decreases the effectiveness of the promotion. Targeted advertising is a practice delivering personalized ads relevant to users as opposed to pushing same ads to all. Personalization for customer targeting across different channels and devices have become a great challenge and seen a huge increase in R\&D expense by firms seeking marketing efficiency \cite{kannan2017digital}. According to studies of TM advertising, targeted advertising gains 115\% more business traffic a year and the targeted consumers have higher brand awareness than random viewers \cite{hu2007demographic}. Targeted advertising is based on information of individual users like geographical location, behavioral variables, and demographic attributes. In this paper, we are interested in demographic attributes (e.g., gender, age, income, etc.) of website audience as they play a key role in providing personalized services. The data used in this paper abides by the standards under U.S. and EU privacy laws and is blind to personally identifiable information (PII), including names, email addresses, phone numbers, financial information, etc.

Traditional methods for determining demographic attributes are mainly based on panel data. The demographic tendency of websites is estimated statistically by known demographic information of panel members who have visited those sites. This approach is reliable for websites visited by a large number of panel members but may lead to a bias for those sites not covered by a sufficient number of panel members. Besides, panel data may be difficult and expensive to obtain. Machine learning approaches construct the relationship between website audience's demographics and their online features (e.g., web content, web category, search keywords, etc.) by building a model on a subset of audiences with known demographics. Then the model can be used to predict other web users' demographic attributes. 
Various other online data has been used to build prediction models, among which social network data is a popular choice. Facebook and twitter profiles have been proven to be very effective in determining the users' personality and demographics \cite{bachrach2012personality, golbeck2011predicting, haenlein2013social, quercia2011our, volkova2015predicting, volkova2016mining}. However, these models require to have access to extensive information about people's social network, which may step into the sensitive area of privacy. 

\begin{figure}
    \centering
    \includegraphics[width=10cm]{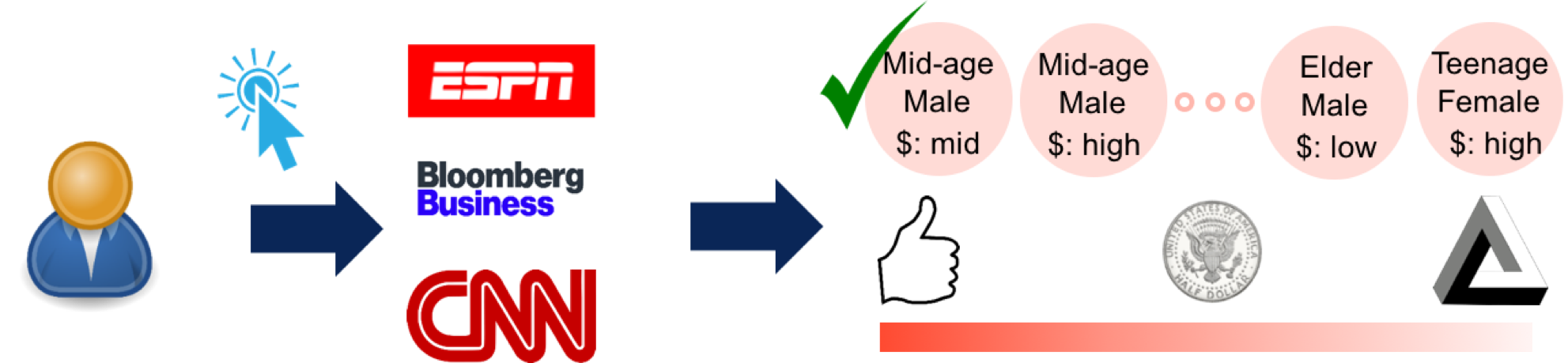}
    \caption{General view of predicting an audience's demographics based on browsing history}
    \label{general_view}
\end{figure}

In this paper, we also take advantage of machine learning approach to predict audiences' demographic attributes.
The proposed method constructs a model to build the relationship between users' browsing history and their demographic attributes. The general view of the method is depicted in Figure \ref{general_view}. The user's identification information can be easily hashed and obfuscated, resulting being PII-compliant. The various information contained in the browsing history has been employed to do the prediction job in literature. The authors of \cite{baglioni2003preprocessing} used web-category information, users' browsing time, browser type and other information to predict personal attributes. In \cite{de2010predicting}, the authors employed more features by including visit frequency, visit duration, web ID, etc. Search keywords entered by users were employed to do the prediction job in \cite{jones2007know,murray2000inferring}. However, such information is usually not available for most users, resulting in a weak generalization of the trained models. 
The most robust method may be content-based methods \cite{hu2007demographic, kabbur2010content}. The web content is always available as long as the website is still there. The proposed model in this paper is also purely based on web content. The only browsing history information needed in our method is hashed users ids together with their browsed domain-level URLs.

Under the proposed framework, firstly we have built a web crawler to retrieve the content of the websites visited by audiences, and represent the content as website feature vectors. Exhaustive experiments have been done to determine the crawling rule. Secondly, we aggregate the vectors of websites that audiences have visited and arrive at feature vectors representing the audiences. Feature representations of audiences are the most important step of the entire framework because existing literature demonstrates that even a simple classifier can achieve good performance as long as it takes in good features as input. At last, a classification model is exploited to predict the audiences' demographic attributes. The support vector machine \cite{cortes1995support} is used as the top classifier in our method because it outperforms other baseline classifiers. Our new model proposes an innovative method to obtain feature vectors representing audiences. The new feature representation method takes advantage of word embedding technique and is easy-to-implement and effective.

Word embedding is a technique representing words with low-dimensional vectors. Word embedding has drawn great attention in recent years because it can capture both semantic and syntactic information. In the vector space, words with similar semantics lie close to each other, for example, \emph{vec(American)} is closer to \emph{vec(France)} than to \emph{vec(Bread)}. Vector representations of words can even preserve the semantic relationship. Word embedding turns out to be quite effective in many natural language processing applications such as parsing \cite{socher2011parsing}, tagging \cite{huang2012improving}, name entity recognition \cite{collobert2008unified}, document summarization \cite{zhang2016extractive,zhang2015multi,zhang2016multiview}, sentiment analysis \cite{er2016attention, zhang2016sentiment} and machine translation \cite{zou2013bilingual}. With advances in deep learning techniques, word vector representation has become a common practice for natural language processing tasks.

Website feature representation in this paper is the same as document representation in natural language processing community. Traditional research works in document representation without using word embedding usually employ the bag-of-words model which uses occurrence frequency of words as features. They face the problem of high dimensionality and sparsity and losing word order of sentences \cite{bengio2003neural,mikolov2013distributed}. To address these issues, most recent works exploit deep learning methods together with word embedding technique and leverage neural networks to construct non-linear interactions between words so as to capture the word order information. The usually used deep learning models include convolutional neural network \cite{er2016attention, kalchbrenner2014convolutional, kim2014convolutional, lecun1998gradient}, recurrent neural networks \cite{funahashi1993approximation,schuster1997bidirectional}, recursive neural networks \cite{socher2011parsing,  socher2013recursive}, paragraph vector \cite{le2014distributed}, etc. The problems with deep learning models are that they are very sophisticated, have a lot of parameters to tune, and are time-consuming to train. 

Considering the specific characteristics of the task in this paper, we propose a simple but effective model to represent website features. As the crawled content of websites is constituted by fragmented information, the word order information may not be very significant. The simplest document representation method based on word embedding is taking the sum or average of all the vectors of words contained in the document if the word order is not important. Our approach is developed based on this simple idea but incorporates a term weighting function based on term frequency–inverse document frequency (tf-idf) formula. With the tf-idf weighting scheme, the document representation vector not only leverages on word co-occurrence probabilities but also takes into consideration the information about the general distribution of terms appearing in different documents. The idea is inspired by \cite{paltoglou2010study}, in which the authors adopted variants of original tf-idf to represent documents for sentiment analysis and achieved significant improvement in classification accuracy. Instead of using the tf-idf vectors to represent documents, our method takes them as weights of word vectors\footnote{Word embedding and word vector appear alternatively in this paper. They have the same meaning.}. The proposed method has both the power of word embedding capturing semantic and syntactic information as well as, the statistical nature of tf-idf. This intuitive representation approach is unsupervised, requiring no human annotations. It is computationally efficient compared with the neural network language models. Experimental results demonstrate that the new method outperforms the baseline document representation methods and achieves even better performance than sophisticated deep learning methods. The main contribution of this paper is summarized as follows:
\begin{itemize}
\item A simple but effective unsupervised document representation method is proposed for targeted advertising. The method takes advantage of pre-trained word embedding and variants of tf-idf weighting scheme.
\item Exhaustive experiments have been done to determine an internal protocol to crawl web content.
\item Two benchmark datasets are created for demographic prediction task.
\end{itemize}

This paper is organized as follows: Section \ref{S:2} gives a brief review of related works. In Section \ref{S:3}, the proposed methodology is described in details. The datasets are introduced in Section \ref{S:4}. Section \ref{S:5} demonstrates the prediction performance of the proposed method. Conclusions are drawn in Section \ref{S:6}.

\section{Related Works}
\label{S:2}
Many researchers have explored to predict audiences' demographics based on browsing behavior. Baglioni \textit{et al.} \cite{baglioni2003preprocessing} predicted gender of users by analyzing users' navigational behavior. They also analyzed the hierarchical ontology design of URLs visited and employed other click-through data (e.g., timestamps, browser type, etc.) as features. The decision tree was used to predict the attribute but only achieved slightly better accuracy than the random guess. Similarly, the authors of \cite{de2010predicting} used more click-through data (e.g. click frequency, browsing duration, etc.) to build features. They used the random forest to predict demographic attributes including gender, age, occupation, and education. The performance can hardly be named satisfactory as well. Jones \textit{et al.} \cite{jones2007know} exploited audiences' query logs to build feature vectors and used support vector machine (SVM) to predict gender and age. They achieved very good prediction performance on both attributes. 
In \cite{murray2000inferring}, the authors used both search keywords and accessed web-pages as input features, and mapped them to a reduced vector space using latent semantic analysis (LSA). Then the reduced vectors are fed into a neural network-based model to predict attributes of unknown audiences. A common characteristic of all the aforementioned methods is that they all used audience-specific data. However, such information is usually not available for most audiences. Therefore, these models may not be generalizable.
Goel \textit{et al.} \cite{goel2012does} used web domains themselves as features to predict five demographic attributes. Each user is represented as a sparse vector whose elements are binary features (e.g., 1 for visited and 0 for not visited). The SVM classifier was trained on a huge panel data and achieved good prediction performance. However, due to the large number of web domains not included the model's prediction capability as it scales by is doubtful. 
The literature so far demonstrates that content-based methods may be the most robust and effective method among all the approaches. 
In \cite{hu2007demographic}, Hu \textit{et al.} used both content-based features and web-category-based features to complement each other, but only achieved a slight performance improvement compared with using content-based features alone. They first predicted the demographic tendency of web-pages and then used a Bayesian network to derive representation vectors for audiences. In \cite{kabbur2010content}, only web content is used and the prediction task is completed by regression-based approaches. The web content in the above two papers is represented by bag-of-words model and standard tf-idf term weighting scheme respectively. The new method proposed in this paper is also only based on website content. 

\section{Methodology}
\label{S:3}

\begin{figure}
    \centering
    \includegraphics[width=12cm]{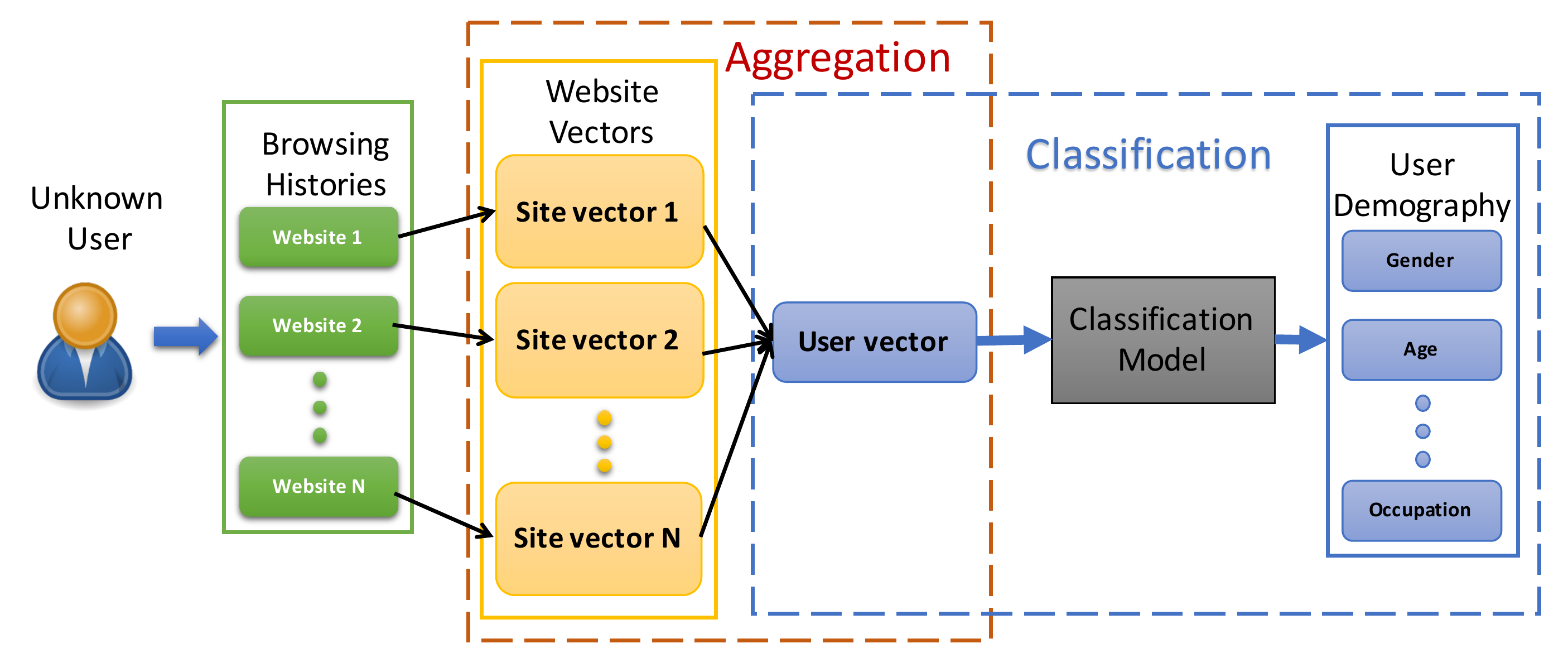}
    \caption{The flow chart that explains the general methodology}
    \label{Flow_methodology}
\end{figure}

In our framework, we first represent the websites browsed by an audience as vectors and then aggregate the website vectors to derive a vector representing the audience. At last, a supervised classification model is used to predict the audience's demographic attributes. The flowchart of our framework is depicted in Figure \ref{Flow_methodology}.
The details of the proposed methodology are discussed in the remainder of this section. The website representation approach is introduced in Section \ref{S:3.2}. Three aggregation methods are discussed in Section \ref{S:3.3}. The classifier is introduced in \ref{S:3.4}.

\subsection{Problem Formulation}
\label{S:3.1}
Before introducing the proposed methodology, we define the problem in this section. A web user's browsing history is a set of websites he or she has visited. Suppose there are totally M users $U=\{u_1, u_2, ..., u_i, ..., u_M\}$ and N websites $S=\{s_1, ..., s_j, ..., s_N\}$. Then the browsing data of all the users can be represented by an adjacency matrix $R$, where the element $r_{ij}$ denotes the weight from user $u_i$ to website $s_j$. The weight is deemed as the visit frequency of user $u_i$ on website $s_j$. Given the demographic attributes $Y=\{y_1, y_2, ..., y_M\}$ of some users are known, the problem is to train a general model ($\mathfrak{M}: U \sim Y$) on the known users and use the trained model to predict demographic attributes of unknown users.

\subsection{Website Representation}
\label{S:3.2}
The first step of our methodology is to represent websites browsed by web users as site vectors. Only content-based features are used in our model. 
In this paper, we propose a new method by taking a powerful technique named word embedding. Variants of the tf-idf weighting scheme are employed as weights of word vectors to help improve the representation power.

\subsubsection{Word Embedding}
\label{S:3.2.1}
There are $M$ websites which need to be represented as vectors in our problem setting. Each website $s_j$ contains a set of words. 
Recent research results have demonstrated that word vectors are more effective in representing words than traditional bag-of-words method. A word can be represented by a dense vector as follows:
\begin{equation}\label{word_embedding}
v = Lx
\end{equation}
where $x \in \mathbb{R}^{V}$ is a one-hot vector where the position that the word appears is one while the other positions are zeros, $L \in \mathbb{R}^{d \times V}$ is a word representation matrix, in which the $i$th column is the vector representation of the $i$th word in the vocabulary, and $V$ is the vocabulary size. 

Word vectors are mostly learned through neural language models \cite{bengio2003neural,mikolov2013efficient,mikolov2013distributed, pennington2014glove,turian2010word}. In neural language models, each word is represented by a dense vector and can be predicted by or used to predict its context representations. However, we can easily adopt off-the-shelf word embedding matrices without spending time and efforts to train our own word vectors. Previous research demonstrated some pre-trained word vectors are universal representations for various tasks and can make good use of semantic and grammatical associations of words. \textit{Word2vec} \cite{mikolov2013distributed} and \textit{GloVe} \cite{pennington2014glove} are two most widely used pre-trained word embedding matrices. Previous research results demonstrate that different performance may result from using either matrix on different tasks, but with slight differences. In this paper, we use \emph{word2vec} \footnote{https://code.google.com/p/word2vec} embedding. 
The \emph{word2vec} vectors are trained on 100 billion words from the Google News by using the Skip-gram method and  maximizing the average log probability of all the words as follows:
\begin{eqnarray}\label{skip_gram}
\frac{1}{T}\sum_{t=1}^{T}\sum_{-c \leq j \leq c, j \neq 0} logp(w_{t+j}|w_t) \\
p(w_O|w_I) = \frac{exp(v_{w_O}'^{T}v_{w_I})}{\sum_{w=1}^{V}exp(v_w'^Tv_{w_I})}
\end{eqnarray}
where $c$ is the context window size, $w_t$ is the center word, $v_w$ and $v_w'$ are the ``input'' and ``output'' vector representations of $w$. More details can be found in \cite{mikolov2013distributed}.

Suppose website $s_j$ is constituted by $K$ words. As every word is represented as a $d$-dimensional vector, the website can be represented as a dense matrix as $V_i = \{v_1,v_2,\dots,v_K\} \in \mathbb{R}^{d \times K}$. The simplest way to reduce the site matrix to a site vector is to take the sum or the average of all the word vectors. Some recent works fulfill the document (website\footnote{A website is regarded as a document in this paper.}) representation task by employing neural networks to model non-linear interactions between words. Our proposed method follows the simple summation fashion but exploits variants of tf-idf weighting scheme to capture the distribution information of words amongst documents. It remains to be very simple and intuitive while our experiment results demonstrate that it even outperforms the sophisticated neural network approaches.

\begin{table}[]
\centering
\caption{Variants of tf weight}
\label{tf_variant}
\begin{tabular}{|c|c|}
\hline
Weighting scheme         & TF         \\ 
\hhline{|=|=|}
default (d) & $f_{t,d}$ \\
\hline
binary (b) & $\left\{\begin{matrix}
1,&f_{t,d}>0 \\ 
0, & otherwise
\end{matrix}\right.$ \\
\hline
logarithm (l) & $1+log(f_{t,d})$ \\
\hline
augmented (a) & $0.5+0.5\cdot \frac{f_{t,d}}{max\{f_{t',d}: t' \in d\}}$ \\
\hline
\end{tabular}
\end{table}

\begin{table}[]
\centering
\caption{Variants of idf weight}
\label{idf_variant}
\begin{tabular}{|c|c|}
\hline
Weighting scheme         & IDF         \\ 
\hhline{|=|=|}
unary (u) & 1 \\
\hline
default (d) & $log(\frac{N}{n_t})$ \\
\hline
smoothed (s) & $1+log(\frac{N}{1+n_t})$ \\
\hline
\end{tabular}
\end{table}
  
\subsubsection{Tf-idf Weighting Scheme}
\label{S:3.2.2}
Tf-idf, short for ``term frequency-inverse document frequency'', is used to reflect the contribution of a word to a document in a corpus. It is widely used in information retrieval and text mining. Typically, tf-idf consists of two components, namely the term frequency (tf) and the inverse document frequency (idf). The two terms are calculated as:
\begin{align}
tf(t,d) &= f_{t,d} \\
idf(t,D) &= log(N/n_t)
\end{align}
where the term $f_{t,d}$ is the raw frequency of term $t$ in a document $d$, $N$ is the total number of documents in the corpus $D$, and $n_t$ is the number of documents with term $t$ in it. And then the weight of term $t$ in the document $d$ is just the product of the two components:
\begin{equation}
w_{t, d} = tf(t,d) * idf(t, D)
\end{equation}

The weight increases proportionally to the term frequency in the given document and is offset by the frequency that the term appears in the entire corpus. This helps to filter the common terms (e.g., the, I, you, etc.) which are actually non-relevant to the specific document. 

Many variations of the tf–idf weighting scheme have been proposed for various applications. Some widely used variants of tf and idf weights \cite{manning2008scoring} are listed in Table \ref{tf_variant} and Table \ref{idf_variant}.  The augmented term frequency is able to prevent a bias towards longer documents.  The smoothed inverse document frequency is proposed in case that $n_t=0$.  

A two-letter notation scheme is used to denote the variants of tf-idf weighting scheme in the following sections according to the notations in Table \ref{tf_variant} and Table \ref{idf_variant}. The first letter denotes the term frequency factor and the second the inverse document frequency. The classic tf-idf weighting scheme is denoted as \textit{dd}. The experiment results demonstrated that the \textit{ad} scheme results in the best information representing power. The weight of term $t$ in a document $d$ under the \textit{ad} scheme is given as follows:
\begin{equation}
w_{t, d}^{ad} = \Big(0.5+0.5\cdot \frac{f_{t,d}}{max\{f_{t',d}: t' \in d\}})\Big) * log(\frac{N}{n_t})
\end{equation}

\subsubsection{Site Vector}
\label{S:3.2.3}
Following the assumption that the website $s_j$ contains $K$ words, the  website can be represented by a dense matrix $V_j = \{v_1,v_2,\dots,v_K\} \in \mathbb{R}^{d \times K}$ by using pre-trained word vectors. The term $d$ is the dimension of pre-trained word vectors. By taking advantage of the tf-idf weighting scheme introduced in Section \ref{S:3.2.2}, we can obtain the weights of the $K$ terms in the website $s_j$. Let's denote the weights of the $K$ terms as a vector $W_{s_j}=\{w_{1, s_j}, w_{2, s_j}, \dots, w_{K, s_j}\}$. Then we can get the site vector using the following equation:
\begin{equation}
s_j =  W_{s_j}V_j^T \in \mathbb{R}^{d}
\end{equation}
Therefore, the website is represented by a low-dimensional vector.

\subsection{Aggregation}
\label{S:3.3}
After representing each website as a vector, an audience's vector can be obtained by aggregating the vectors of all the websites browsed by him or her. This can be easily done by taking a weighted average of all the website vectors by using the weight matrix $R \in \mathbb{R}^{M \times N}$ which gives the visit frequency of each user to all the websites $S=\{s_1, s_2, ..., s_N\} \in \mathbb{R}^{d \times N}$.
\begin{equation}
\label{WA}
u_i^{WA} =  g(R_{i\cdot})S^T \in \mathbb{R}^{d}
\end{equation}
where $g(\cdot)$ is a function to rescale values in a vector so that they sum up to be $1$. The weighted average method faces a problem that some websites have much higher visit frequency thus weights than others. To address the problem, we narrow down the weight differences by mapping visit frequency to log space as follows:
\begin{equation}
\label{LA}
u_i^{LA} =  g(log(R_{i\cdot}))S \in \mathbb{R}^{d}
\end{equation}
The implementation can effectively eliminate the influences of large weight difference. For example, suppose the visit frequency to sites $\{s_1, s_2\}$ are $\{10, 1000\}$, the weight ratio would be 1:100 using Equation \ref{WA} and 1:3 using Equation \ref{LA}. 

Another even simpler aggregation method is to take the simple average of the browsed website vectors by ignoring the user wights on websites as follows:
\begin{equation}
u_i^{SA} =  \widetilde{R}_{i\cdot}S \in \mathbb{R}^{d}
\end{equation}
Where $\widetilde{R}$ is derived from $R$ with the following formula:
\begin{equation}
\widetilde{R}_{ij}=\left\{\begin{matrix}
1,&R_{ij}>0 \\ 
0, & R_{ij}=0
\end{matrix}\right.
\end{equation}

\subsection{Classification}
\label{S:3.4}
We use support vector machine (SVM) as our classifier. However, we have still done several experiments comparing the learning performance of SVM with several other baseline machine learning algorithms, i.e., logistic regression, neural network and random forest. The results demonstrate that SVM achieves similar performance as neural network and outperforms logistic regression and random forest.

\section{Datasets}
\label{S:4}

\subsection{Demographic Attributes}
There are many important demographic attributes associated with a person, e.g. age, gender, race, household income and so on. In this paper, we focus on gender and age, because they are easily and legally accessible. However, the predictive model that we developed can be easily applied to any other types of audience classification as long as the attributes are allowed to be used by law. The gender attribute is categorized into male and female, while the age attribute is broken down into four groups, i.e., Teenage (\textless 18 years), Young (18–34 years), Mid-age (35–49 years) and Elder (50+ years). The breakdown method closely corresponds to groups that are of interest to advertising agencies. 

\subsection{Data Preparation}
In our experiments, we use panelist data for training our learning model, and the trained model is applied on cookie browsing data for prediction. 
The panel data used contains audiences' browsing log data in May 2015. Each record of the data consists of user id, web-page clicked, visit frequency of that web-page, and gender/age information. The data contains 175640 distinct users and 2476338 unique websites. The websites with very low traffic (visit frequency of all users less than 100) are removed as the content of such websites falls out of the scope of most users. If a website is visited by a user by less than five times, it is taken away from the list of visited websites of that user, since it is not reasonable to refer a user's attributes using websites that he or she randomly visits. Then the websites that cannot be crawled are filtered out. We also filter the websites whose crawled content has less than 10 words because such few content can hardly convey useful information. The number of websites clicked by a user is an important variable for user's demography prediction, because it is not reasonable to predict a user's demographic information based on very few websites. In this paper, we choose two threshold values for a minimum number of websites visited, namely 20 and 100 to investigate the impact. Two datasets, namely \textit{Demo20} and \textit{Demo100}, are generated by filtering out the users who visit less than the corresponding number of websites. After all the processing, \textit{Demo20} has 70868 distinct users and 3499537 entries in sum, while \textit{Demo100} has 4742 distinct users and a total of 667019 entries. The gender and age distribution of the two datasets are shown in Table \ref{distribution_breakdown_demo20} and Table \ref{distribution_breakdown_demo100}.  

\begin{table}[]
\parbox{.45\linewidth}{
\centering
\caption{Data distribution over gender and age of Demo20}
\label{distribution_breakdown_demo20}
\begin{tabular}{|c|l|l|l|}
\hline
\multirow{2}{*}{Age} & \multicolumn{2}{c|}{Gender} & \multicolumn{1}{c|}{\multirow{2}{*}{Total}} \\ \cline{2-3}
                     & Male         & Female       & \multicolumn{1}{c|}{}                       \\ \hline
Teenage              & 3.5\%        & 3.1\%        & 6.6\%                                       \\ \hline
Young                & 21.9\%       & 16.7\%       & 38.7\%                                      \\ \hline
Mid-age              & 16.7\%       & 17.2\%       & 33.9\%                                      \\ \hline
Elder                & 10.3\%       & 10.5\%       & 20.8\%                                      \\ \hline
Total                & 52.5\%       & 47.5\%       & 100\%                                       \\ \hline
\end{tabular}}
\hfill
\parbox{.45\linewidth}{
\centering
\caption{Data distribution over gender and age of Demo100}
\label{distribution_breakdown_demo100}
\begin{tabular}{|c|l|l|l|}
\hline
\multirow{2}{*}{Age} & \multicolumn{2}{c|}{Gender} & \multicolumn{1}{c|}{\multirow{2}{*}{Total}} \\ \cline{2-3}
                     & Male         & Female       & \multicolumn{1}{c|}{}                       \\ \hline
Teenage              & 1.6\%        & 1.3\%        & 2.9\%                                       \\ \hline
Young                & 18.3\%       & 13.2\%       & 31.4\%                                      \\ \hline
Mid-age              & 18.7\%       & 18.2\%       & 36.9\%                                      \\ \hline
Elder                & 16.4\%       & 12.4\%       & 28.8\%                                      \\ \hline
Total                & 54.9\%       & 45.1\%       & 100\%                                       \\ \hline
\end{tabular}}
\end{table}

\subsection{Web Crawling}
Web crawling of websites is necessary in our task, in order to extract key information from a web page. As websites are now designed to appear in different formats and styles to appeal to a diverse audience, and different web platforms (eg. mobile and computer), it is necessary for us to identify an internal protocol to extract web content, which is ubiquitous to most websites, in a generalized manner. In our implementation, the HTML tag-based method is used to crawl the websites. 

Specifically, the title tag $<$title$>$, headline tags $<$h1-h6$>$, paragraph tags $<$p$>$, link tags $<$a$>$, and image tags $<$img$>$ are chosen and tested on the gender attribute of a website demography tendency dataset. This dataset contains the top 2000 visited websites in the USA, with the visitors' demographic tendency for each website. To build our protocol, we tested a combination of the tag information collected in our experiments. Concretely, we want to understand if a certain (combination of) tag information will be more reliable and useful to succinctly describe a website. Intuitively, we believe that the $<$title$>$ tag conveys very important information about a website, therefore, it is always included in our protocol. As these five HTML tags do not encompass the entire website's content, we have also crawled the visible text in each website. Although the visible text contains more information for each website, it also includes more noise for our learning model. This is demonstrated in our experimental results. 

\begin{figure}
  \centering
  \includegraphics[width=0.5\textwidth]{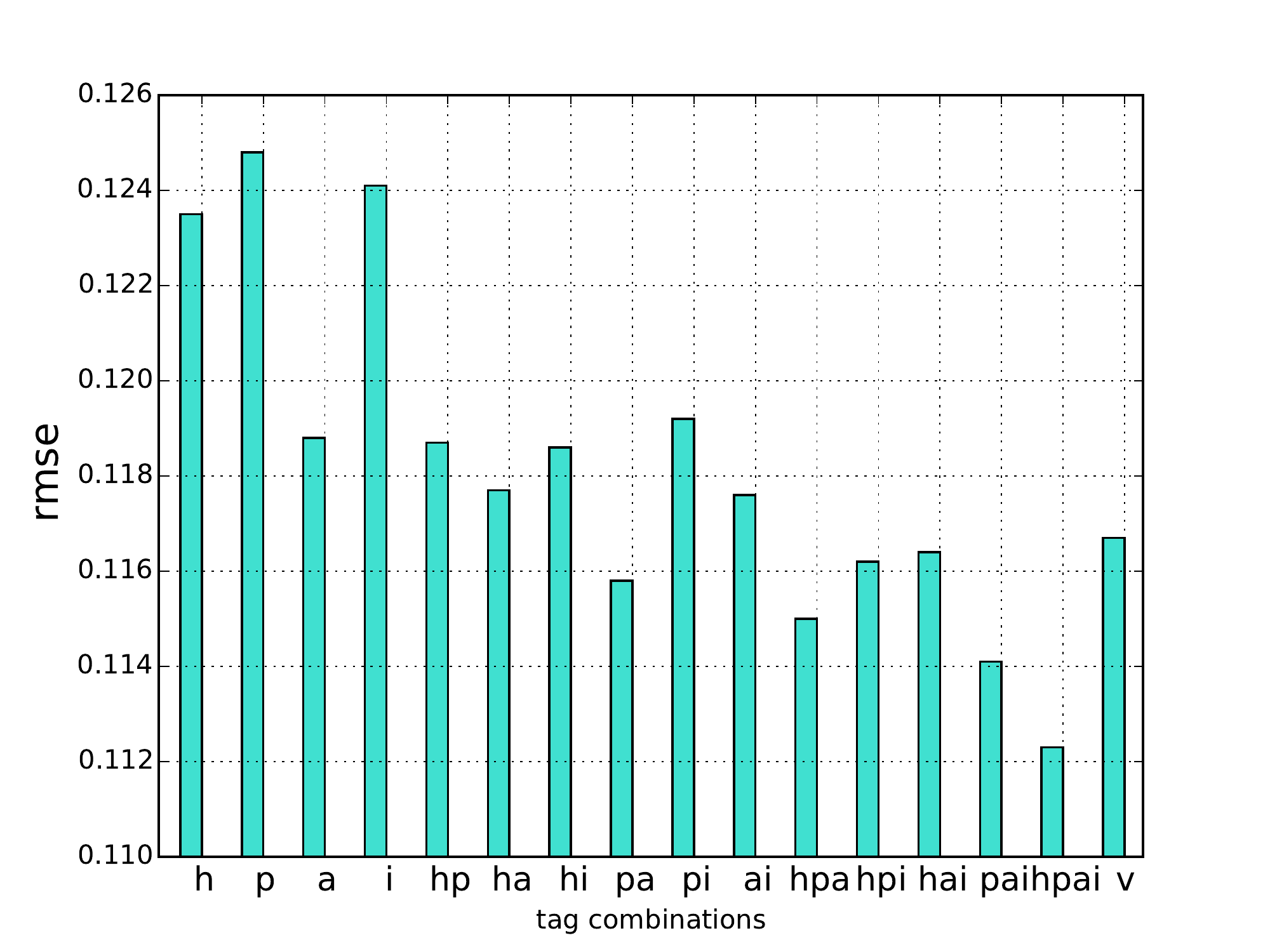}
  \caption{RMSE results by tag combinations on website demography tendency prediction. The terms $h$, $p$, $a$, $i$, $v$ are used to represent headlines, paragraphs, link words, image descriptions, and visible texts respectively for brevity. Tag combination $xy$ means content associated with the corresponding tags $x$ and $y$ are used. For instance, $hp$ stands for using both headlines and paragraph content.}
\label{tag_comb_perf}
\end{figure}

The website demography tendency dataset uses continuous scores to indicate visitors' demographic tendency (e.g., 0.45 for male and 0.55 for female). As such, we employ a regression model named $\epsilon$-support vector regression ($\epsilon$-SVR) \cite{smola2004tutorial}, which is an extension of SVM, to predict the tendency scores for a given website. The prediction root mean square error (RMSE) results of all the tag combinations are demonstrated in Figure \ref{tag_comb_perf}. We observe that the combination $hpai$ achieves the smallest RMSE, while the inclusion of all visible text does not improve the performance due to the presence of additional noise.
Therefore, the combination of $hpai$ is employed to crawl website content in our model.

\section{Experiment Analysis}
\label{S:5}
In this section, we evaluate the performance of the proposed model on the two audience demography datasets, namely \textit{Demo20} and \textit{Demo100}. The impact of different tf-idf weighting schemes on website representation is evaluated in Section \ref{S:5.2}. The three aggregation methods are compared in Section \ref{S:5.3}. The prediction performance using different classifiers are shown in Section \ref{S:5.4}. Section \ref{S:5.5} demonstrates the power of new proposed website representation method by comparing with existing baseline and state-of-art methods. One more experiment is done in Section \ref{S:5.6} to demonstrate that our aggregation-classification framework is better than regression-aggregation framework used in \cite{hu2007demographic}.

\subsection{Experiment Setup}
\label{S:5.1}
The greatest advantage of the proposed model is that the audience feature representation procedure has no parameters to tune. We directly use the word embedding matrix \textit{word2vec} which has been pre-trained on Google news. The dimension of each word vector is 300. Gensim toolkit \cite{rehurek_lrec} is exploited to obtain the tf-idf weights. The only parameters to tune are the parameters in the SVM classifier. The SVM classifier is implemented using the LinearSVC package in scikit-learn toolkit \cite{scikit-learn}. Ridge norm penalty is used and the penalty parameter $C$ is determined using grid search and cross-validation. The two datasets are split into training and testing datasets by 3:2 ratio. The linear SVC is trained on the training dataset and all the prediction results in the following sections are obtained on the testing dataset.

\subsection{Performance of Different Tf-idf Weighting Schemes}
\label{S:5.2}
Four variants of tf weight and three variants of idf weight are shown in Table \ref{tf_variant} and Table \ref{idf_variant}. Therefore, there are totally 12 different tf-idf weighting schemes. In this section, we compare the demography prediction results on both \textit{Demo20} and \textit{Demo100} using all the 12 weighting schemes. As the comparison results on gender attribute and age attribute are similar, only gender prediction results are shown for conciseness. The log average aggregation implementation is used. 

\begin{figure}
  \centering
  \includegraphics[width=0.6\textwidth]{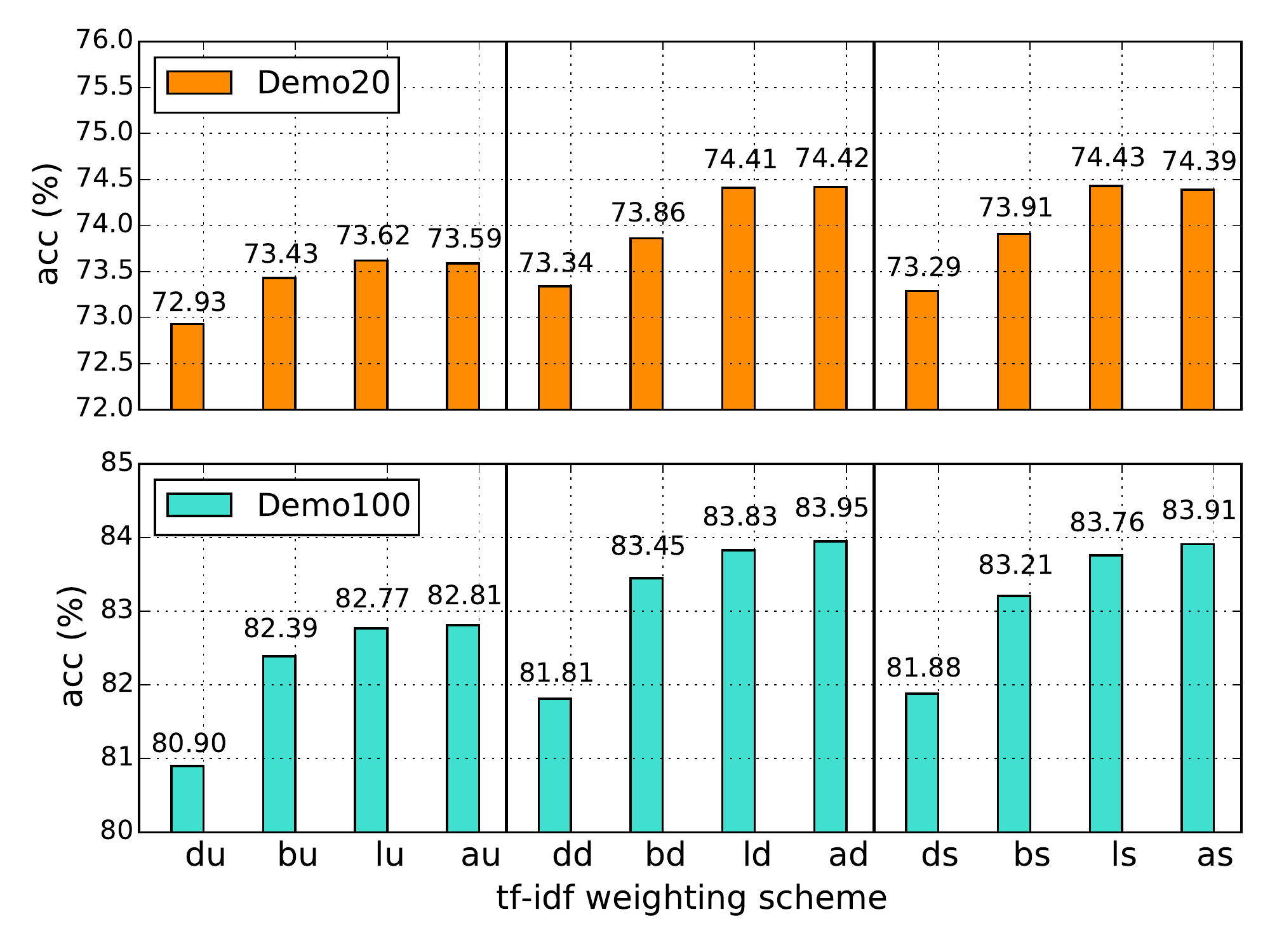}
  \caption{Accuracy comparison results on gender prediction using various tf-idf weighting scheme on \textit{Demo20} and \textit{Demo100} datasets. The first letter in the scheme notation denotes the name of the variant of term frequency weight. The second letter accounts for the variant of inverse document frequency weight.}
\label{tfidf_perf}
\end{figure}

Figure \ref{tfidf_perf} demonstrates the accuracy comparison results on gender prediction using various tf-idf weighting scheme on \textit{Demo20} and \textit{Demo100} datasets. The results on both \textit{Demo20} and \textit{Demo100} show that binary tf weight scheme performs better than the default raw tf weight, e.g., \textit{bd} (83.45\%) vs. \textit{dd} (81.81\%). Furthermore, other variants of tf weight also outperform the raw tf weight. The logarithm and augmented weighting schemes perform similarly, both slightly better than binary features, e.g., \textit{ld} (83.83\%) vs. \textit{ad} (83.95\%). We can conclude that sub-linearly scaled tf weight improves representation capability compared with raw term frequency.

When comparing the idf weighting schemes, we can find that the default idf weight helps improve representation power compared with unary idf weight, e.g., \textit{ad} (83.95\%) vs. \textit{au} (82.81\%). The \textit{du} weighting scheme actually takes the weighted average of vectors of all words on the website, without considering the word distribution information over the corpus.
With the incorporation of idf weighting scheme, the website representation vector not only leverages word co-occurrence probabilities but also takes into consideration the information about the general distribution of terms appearing in different documents. In \cite{paltoglou2010study}, the authors claimed that the idf smoothing greatly improved the classification accuracy on sentiment analysis task. However, our results show that smoothing idf weight does not provide any advantage over the classic idf weight. This is because all the terms appear in the corpus at least once ($n_t >= 1$). In our task, the \textit{ad} weighting scheme achieves the best prediction performance on both \textit{Demo20} and \textit{Demo100}.

\begin{figure}
  \centering
  \includegraphics[width=0.6\textwidth]{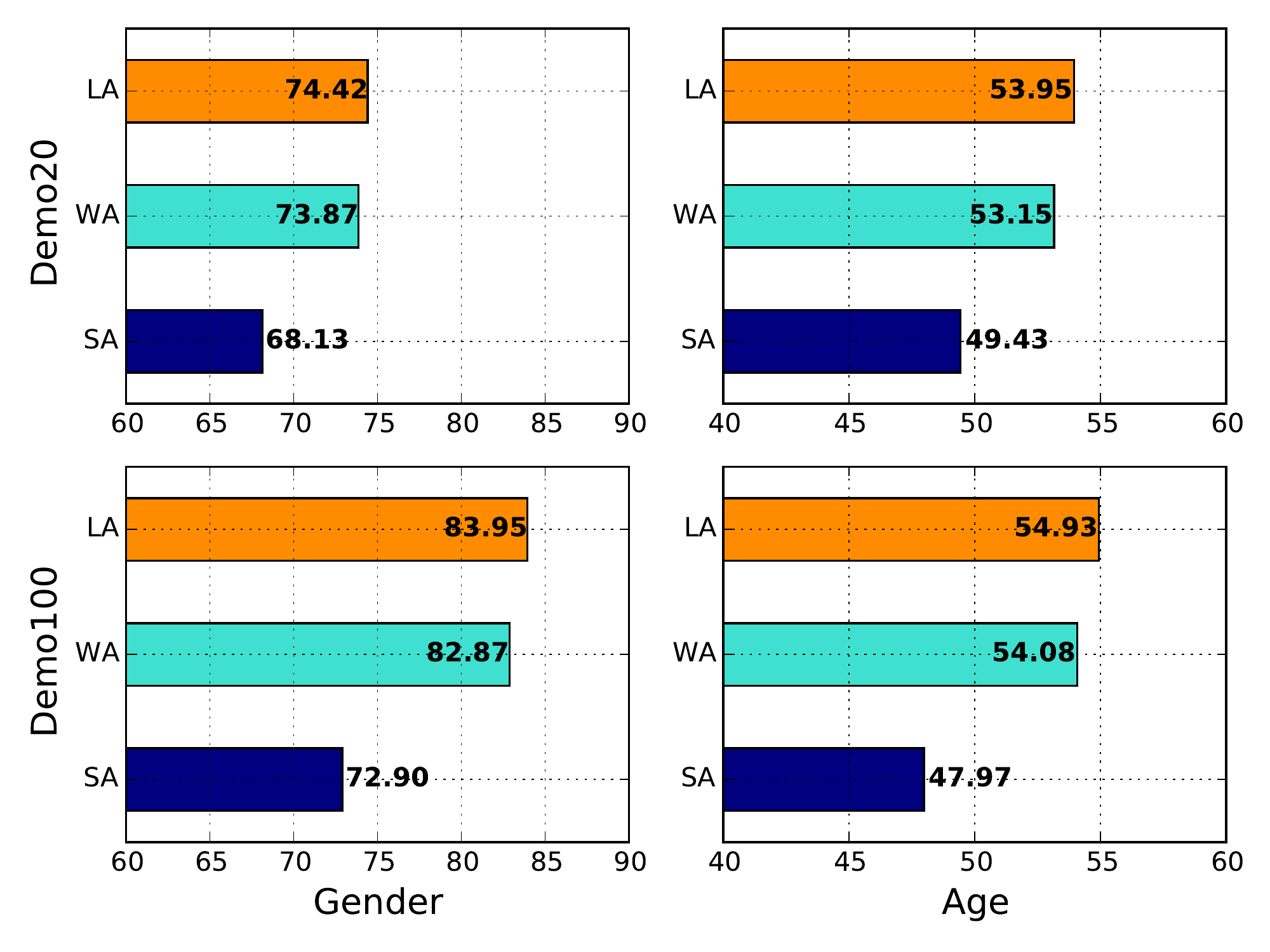}
  \caption{Accuracy comparison results using various aggregation methods. The orange (LA),turquoise (WA), and navy (SA) bars denote log, weighted, and simple average methods respectively. The upper two sub-figures show results on \textit{Demo20} dataset while the bottom figures on \textit{Demo100}. The left two sub-figures are results on gender attribute prediction while the right two on age.}
\label{aggregation_perf}
\end{figure}

\subsection{Performance of Different Aggregation Methods}
\label{S:5.3}
After obtaining the representation vectors for websites, an aggregation step is needed to move from website vectors to audience vectors. In Section \ref{S:3.3}, three aggregation methods are introduced. They are compared in this section. The best tf-idf weighting scheme in our task, namely \textit{ad}, is used in all the experiments in this section.
The comparison results on two datasets for both gender and age attributes are demonstrated in Figure \ref{aggregation_perf}. We can see that the prediction accuracy using weighted average method is much poorer than that using simple average aggregation although the former taking into consideration the web user's visit frequency on each website. This results from the fact that frequency of few websites may be tens of or hundreds of times larger than that of other sites. Thus few most frequently visited websites dominate the final audience representation and contributions of other websites can almost be neglected. The best aggregation method among the three is the log average method. It still leverages on the visit frequency difference, but rescale the weights to a much narrower range so that all the useful websites' information is included in the final audience representation. According to the experimental results in Section \ref{S:5.2} and Section \ref{S:5.3}, the ad tf-idf weighting scheme and log average aggregation method are employed to obtain the final audience vector representation.

\subsection{Performance of Different Classification Models}
\label{S:5.4}
Several popular baseline classifiers are used as the classification model in our framework and their prediction performance on gender attribute is shown in Table \ref{classifier}. All the classifiers have been implemented using scikit-learn toolkit \cite{scikit-learn}. For the neural network, one hidden layer perceptron classifier has been used. Random forest and logistic regression are implemented using base classifiers in the toolkit. Default settings are used for most parameters. The most important hyper-parameters, like regularization strength of logistic regression, the number of maximal features and estimators of the random forest, and the number of hidden nodes of the neural network are determined using grid search and cross-validation. The results in Table \ref{classifier} demonstrate that SVM achieves similar performance as neural network and outperforms logistic regression and random forest. Random forest performs rather poor in this task, which should be expected from the noisy characteristic of web data. The fact that a linear SVM achieves rather satisfactory performance demonstrates that the feature representations of audiences are effective and informative. 

\begin{table}[]
\centering
\caption{Gender prediction accuracy of SVM against other baseline classifiers (\%)}
\label{classifier}
\begin{tabular}{|c|c|c|}
\hline
Models              & \multicolumn{1}{c|}{Demo20} & \multicolumn{1}{c|}{Demo100} \\ \hline
Logistic Regression &            73.39            &             82.53            \\ \hline
Random Forest       &            67.63            &            77.60            \\ \hline
Neural Network      &            74.28            &              83.99            \\ \hline
SVM                 &            74.42            &              83.95            \\ \hline
\end{tabular}
\end{table}

\subsection{Comparison of Website Representation Methods}
\label{S:5.5}
One of the main contributions of the paper is our proposed website representation approach. Our method is purely based on content and does not require any other audience-specific input data. To demonstrate the representation power of the proposed method. We compare it with some other website representation methods. The comparison methods are described as follows:
\begin{itemize}
\item {\bf Category-based method}: represents websites using category-based features. Websites are spanned over 460 two level categories from Open Directory Project (ODP\footnote{http://dmoz.org}). ODP is the largest, most comprehensive human-edited directory of the Web. Each website is represented by a 460-dimensional vector with each value indicating the possibility of belonging to each category. The possibility values are obtained by comparing the web content with keyword library built for each category.
\item {\bf Latent Semantic Analysis (LSA)}: is a widely used topic modeling method which is able to find the relationships between terms and documents. Gensim toolkit \cite{rehurek_lrec} is used to implement LSA and the number of topics is set as 300.
\item {\bf Recurrent Neural Network (RNN)}: is a powerful deep learning model which is able to construct non-linear interactions between words. It takes in word vectors in a website one by one and outputs the vector representation of the website. The model is trained using demographic tendency scores of websites which are propagated from attributes of audiences who have browsed the websites. It is implemented using keras\footnote{http://keras.io}.
\item {\bf TF-IDF}: uses tf-idf weights directly as feature vectors of websites without using word vectors. We also use gensim toolkit to obtain tf-idf representations and the default tf-idf is used.
\end{itemize}

\begin{table}[]
\centering
\caption{Comparison of the percentage (\%) accuracy of different websites representation methods}
\label{sitevec_comparison}
\begin{tabular}{|c|c|c|c|c|}
\hline
\multirow{2}{*}{\begin{tabular}[c]{@{}c@{}}Representation\\ methods\end{tabular}} & \multicolumn{2}{c|}{Demo20} & \multicolumn{2}{c|}{Demo100} \\ \cline{2-5} 
                                                                                  & Gender         & Age        & Gender         & Age         \\ \hline
Category-based                                                               &                 71.71&      53.65      &             79.34   &     53.03        \\ \hline
LSI                                                                     &                
71.93&        53.49    &      79.81        &     53.55        \\ \hline
RNN                   &                                                 68.35&  49.65  &            75.49&         50.71                    \\ \hline
TF-IDF                                              &                70.56&        52.21    &         77.96       &        52.78     \\ \hline
TF-IDF\_word2vec                                &                                                 
\textbf{74.42} &         \textbf{53.95}       &      \textbf{83.95}      &        \textbf{54.93}                  \\ \hline
\end{tabular}
\end{table}

The comparison results are demonstrated in Table \ref{sitevec_comparison}. The last row is the representation method proposed in this paper. The \textit{ad} variant of tf-idf weighting scheme is used. We can see that the new representation method achieves a much better performance than other approaches. The performance of category-based method relies on the completeness and precision of keyword library of each category, which is not easy to build. This may be the reason that category-based method only achieves average performance. LSI has not achieved very satisfactory performance either, demonstrating that topic extraction from messy website content is not an easy task. Comparing TF-IDF and our word embedding method, we can conclude that the incorporation of word vectors help significantly improve the representation power. TF-IDF representations are very sparse and with high dimension, resulting in difficulty in training a good classifier. What is of interest is that the sophisticated deep learning model RNN shows the worst performance. One reason is that website content usually contains a lot of words. RNN faces great challenges to memorize long sequences because of the problems of `vanishing gradient' and `explosive gradient'. Another reason is that the content in websites is usually fragmented, therefore, this greatly limits the use of deep learning within our framework. Furthermore, sophisticated RNN model requires a great amount of time to train while our model is computationally efficient. The comparison results demonstrate the effectiveness and power of the new website representation method.

\subsection{Study on Overall Framework}
\label{S:5.6}
\begin{figure}
  \centering
  \includegraphics[width=0.5\textwidth]{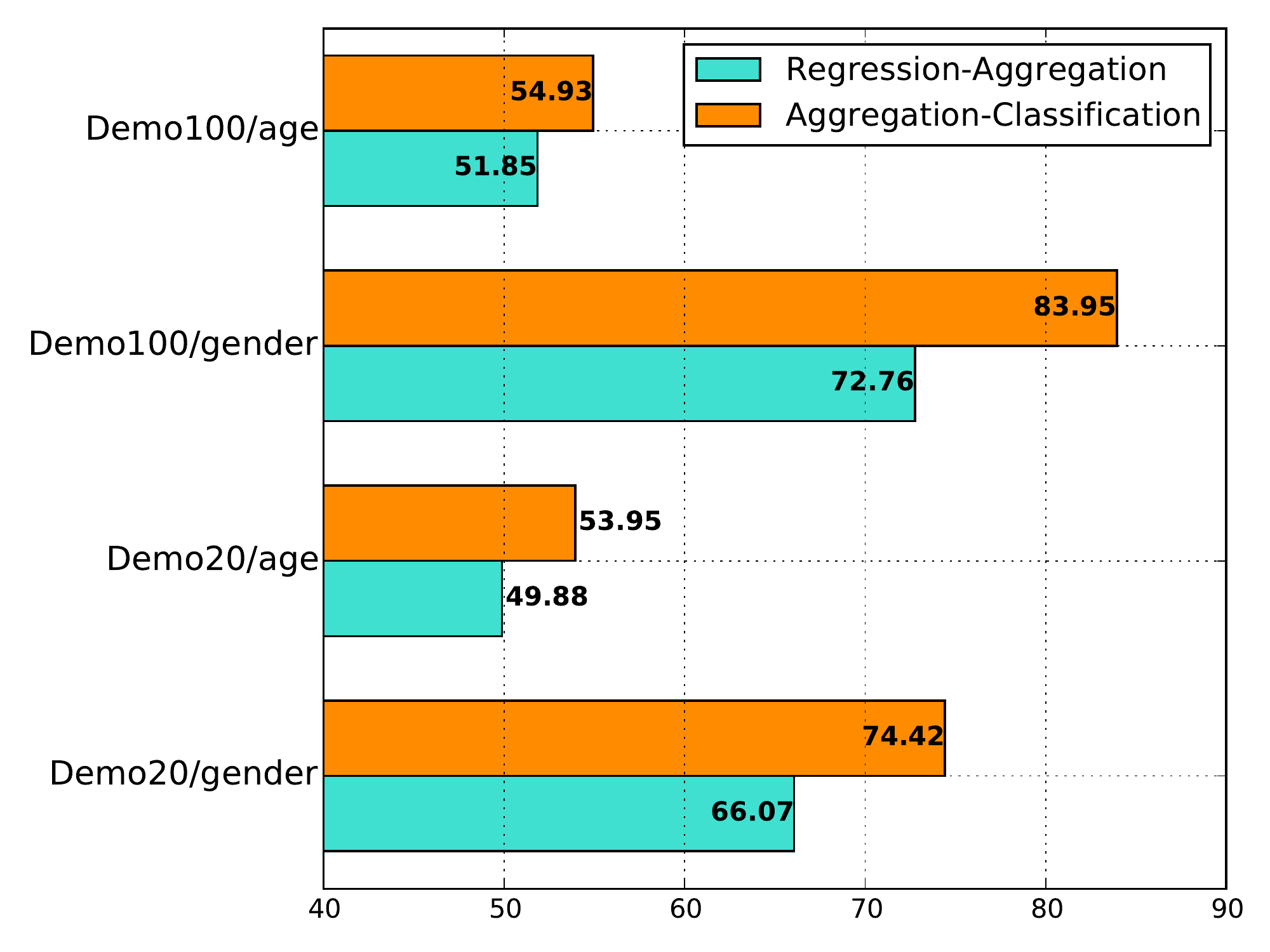}
  \caption{Accuracy comparison results using aggregation-classification and regression-aggregation framework. The upper two are age and gender prediction on \textit{Demo100} dataset, while the bottom two are prediction results on \textit{Demo10}.}
\label{framework_comp}
\end{figure}

In \cite{hu2007demographic}, Hu \textit{et al.} first trained a supervised regression model to predict a web-page’s gender and age tendency. Then, an audience's gender and age are predicted based on the age and gender tendency of the web-pages browsed by the audience within a Bayesian framework. The Bayesian framework assumed that the web-pages visited by the audiences are independent.  Our model does audience attribute prediction after aggregating information of websites browsed by the user. In this section, we compare the two frameworks. The framework in Hu's paper is denoted as regression-aggregation, while our framework as aggregation-classification. The comparison results are depicted in Figure \ref{framework_comp}. We can see that the aggregation-classification framework achieves much higher accuracy than regression-aggregation framework when predicting both gender and age attribute on \textit{Demo20} and \textit{Demo100}. The results indicate the independent assumption is problematic. The fact is that websites visited by the same web user are usually similar or correlated. The experimental results demonstrate that our framework is more reasonable.

\section{Conclusion}
\label{S:6}
Audience classification is of significant importance for targeted advertising.  In this paper, a new method is proposed to estimate on-line audiences' demographic attributes based on their browsing histories, as an example of audience classifications. We first retrieve the content of the websites visited by audiences and employ an innovative website representation method to represent the content as feature vectors. Word embedding with variants of tf-idf weighting scheme turns out to be a simple but effective website representation method. The method is unsupervised and requires no human annotations. Various variants of tf-idf weighting schemes have been tested and it is observed that sub-linearly scaled term frequency weight together with the default inverse document frequency weight can significantly improve the final classification accuracy. After obtaining website vectors, a log average aggregation method is shown to be effective in composing vectors of websites browsed by the audience into a vector representing that audience. The experimental results demonstrate that our new audience feature representation method is more powerful than existing baseline methods, leading to a significant improvement in prediction accuracy. Another contribution of the paper is examining various web crawling rules and identifying the best crawling pattern for audience classification. For future work, the proposed methodology will be tested on other demographic attributes, like income, race, sexual orientation, and hobbies, etc.






\bibliographystyle{model4-names}

\bibliography{main}

\end{document}